\documentclass[sigconf]{acmart}

\renewcommand\footnotetextcopyrightpermission[1]{}
\usepackage[T1]{fontenc}
\IfFileExists{lmodern.sty}{\usepackage{lmodern}}{}
\usepackage{microtype}
\usepackage{hyperref}
\usepackage{url}
\usepackage{graphicx}
\usepackage{amsmath}
\usepackage{booktabs}
\usepackage{tabularx}
\usepackage{array}
\usepackage{fancyvrb}
\usepackage{fvextra}
\usepackage{enumitem}
\usepackage{placeins}
\usepackage{multicol}

\fvset{fontsize=\footnotesize,breaklines,breakanywhere}

\title[DMA Streaming Framework]{The DMA Streaming Framework: Kernel-Level Buffer Orchestration for High-Performance AI Data Paths}
\subtitle{A General-Purpose Buffer Orchestration Layer for High-Performance AI}

\author{Marco Graziano}
\affiliation{%
  \institution{Graziano Labs Corp.}
  \city{Palo Alto}
  \state{CA}
  \country{USA}
}
\email{marco@grazianolabs.com}

\begin{abstract}
AI transport libraries move bytes efficiently, but they commonly assume that buffers are already correctly allocated, placed, shared, registered, and safe under completion and teardown pressure. This paper presents \textit{dmaplane}, a Linux kernel module that makes this missing layer explicit as \textit{buffer orchestration}. dmaplane exposes a stable kernel UAPI via \texttt{/dev/dmaplane} and composes ring-based command channels, DMA buffer lifecycle management, dma-buf export for cross-device sharing, a kernel-space RDMA engine, NUMA-aware allocation and verification, credit-based flow control, low-overhead observability, and GPU memory integration via PCIe BAR pinning. We evaluate orchestration sensitivity with measurements of NUMA cross-node penalties at DRAM scale, completion-safe flow control under sustained RDMA load, and GPU BAR mapping tiers versus \texttt{cudaMemcpy}. We also demonstrate end-to-end disaggregated inference by transferring KV-cache chunks between two machines using RDMA WRITE WITH IMMEDIATE and reconstructing tensor views on the receiver. RDMA measurements use Soft-RoCE; we distinguish measured results from provider-independent properties by construction.
\end{abstract}

\acmConference{Preprint}{2026}{}
\acmBooktitle{}
\acmYear{2026}
\acmDOI{}
\acmISBN{}
\acmPrice{}

\begin{document}
\maketitle

\section{Introduction}

Modern AI systems are limited by host side data movement as often as by accelerator compute. Large language model workloads increasingly rely on point to point transfers between GPUs across hosts, including disaggregated inference where a prefill GPU produces KV cache state and a decode GPU consumes it~\cite{patel2024splitwise,zhong2024distserve}. Transport systems can move bytes efficiently, but they commonly assume that buffers are already allocated on the correct NUMA node, shareable across devices without copies, registered for RDMA, and safe under completion pressure and teardown. These preconditions are not transport. They are buffer orchestration.

This paper presents \textbf{dmaplane}, a Linux kernel module that makes buffer orchestration an explicit kernel layer. dmaplane exposes a stable UAPI at \texttt{/dev/dmaplane} using \texttt{ioctl} and \texttt{mmap} plus optional dma-buf file descriptors, and composes ring based command channels, DMA buffer lifecycle management, dma-buf export for cross device sharing, a kernel space RDMA engine, NUMA aware allocation and verification, credit based flow control for completion safety, low overhead observability, and GPU memory integration via PCIe BAR pinning.

The repo source code is at \url{https://github.com/marcoeg/dmaplane}.

\subsection{Goals and non goals}

\textbf{Goals.} dmaplane has two goals:

\begin{itemize}[leftmargin=*,nosep]
\item Provide a coherent kernel subsystem for buffer lifecycle, placement, sharing, and flow control, integrated with dma-buf, NUMA placement, RDMA memory registration, and GPU BAR pinning constraints.
\item Expose a stable kernel UAPI, based on \texttt{ioctl} and \texttt{mmap} plus optional dma-buf file descriptors, that higher level transports can consume.
\end{itemize}

\textbf{Non goals.} dmaplane is not intended to provide:

\begin{itemize}[leftmargin=*,nosep]
\item Collective communication primitives or an NCCL replacement.
\item A serving framework, including batching, routing, prefix caching, or scheduler integration.
\item Connection management (\texttt{rdma\_cm}), multi-tenant isolation, or production fault recovery.
\end{itemize}

\subsection{Motivating workloads beyond disaggregated inference}

Disaggregated inference~\cite{patel2024splitwise,zhong2024distserve,qin2024mooncake} is a primary motivation because it requires point to point KV cache movement with receiver notification and bounded completion pressure. The same buffer orchestration layer is required in several adjacent workloads.

\textbf{Mixture of Experts inference.} MoE dispatch and combine routes token batches to expert networks on different devices. Activations become fine grained point to point messages. Steady state throughput depends on staging buffer placement, repeated registration cost, and completion safety under bursty traffic.

\textbf{Disaggregated training.} Training pipelines increasingly separate data parallel and model parallel stages from optimizer and checkpoint services. Gradients, optimizer shards, and activation checkpoints move between devices and hosts without a single global collective barrier. Buffer lifecycle, cross device sharing, and teardown safety are required when multiple DMA engines touch the same buffers.

\textbf{RL rollout weight transfer.} Reinforcement learning systems commonly separate actors producing rollouts from learners updating weights. Periodic weight pushes from learners to many actors stress point to multipoint distribution. Orchestration determines whether weights can be staged once, shared, and transmitted efficiently without CPU copies and without completion overflow during fanout bursts.

\textbf{LLM training with weight streaming.} In weight streaming architectures, such as Cerebras weight streaming, host side pipelines stage large weight shards and stream them into the accelerator each step. The limiting factors are sustained bandwidth, locality to the correct NUMA node and PCIe root complex, and backpressure correctness. Buffer orchestration governs placement, sharing, and flow control under sustained streaming.
\subsection{Contributions}

This paper makes four contributions:

\begin{enumerate}
\item A concrete design and implementation of a kernel level buffer orchestration layer with explicit invariants for concurrency, lifecycle, and teardown (Sections 3 and 4).

\item Measured sensitivity results showing where orchestration choices dominate throughput and safety: NUMA placement effects at DRAM scale (Table 4), completion safe flow control under sustained load (Table 3), and GPU BAR mapping tier behavior versus cudaMemcpy (Table 5).

\item An end to end disaggregated inference demonstration that transfers KV cache chunks between two machines using RDMA WRITE WITH IMMEDIATE and reconstructs tensor views on the receiver (Section 5, Table 2).

\item A practical decomposition into independently testable subsystems with reproducibility guidance and appendices for build, deployment, and test mapping (Appendices A through C).
\end{enumerate}

\subsection{Paper organization}

Section 2 summarizes required background and positions dmaplane relative to adjacent systems. Section 3 describes architecture and design invariants. Section 4 presents the core mechanisms in dependency order. Section 5 describes the disaggregated inference demo and protocol. Section 6 evaluates the system and discusses limitations. Section 7 concludes. References follow the main body.
\section{Background and Related Work}

dmaplane relies on existing Linux kernel primitives rather than reimplementing memory management or RDMA transport. This section states only the background required to follow the design and then positions dmaplane relative to neighboring systems.

\subsection{Kernel primitives used by dmaplane}

\textbf{DMA allocation and mapping.} Small control structures often use \texttt{dma\_alloc\_coherent} to obtain coherent memory and a DMA address for a specific device. Large data buffers typically use page backed allocation (\texttt{alloc\_pages} and \texttt{alloc\_pages\_\allowbreak node}) and device specific mapping through scatter gather tables and the DMA mapping API.

\textbf{NUMA placement.} \texttt{alloc\_pages\_node(node, ...)} requests allocation from a specific NUMA node, but can fall back silently to other nodes. Correct placement therefore requires explicit verification after allocation.

\textbf{dma-buf.} The dma-buf framework enables zero copy sharing of DMA capable memory between kernel drivers via an exporter importer contract. A key constraint is that scatter gather tables must be constructed per importer attachment because DMA addresses depend on the importing device and its IOMMU context.

\textbf{Kernel verbs RDMA.} The InfiniBand verbs API in \texttt{ib\_core} provides a provider independent interface for RDMA~\cite{corbet2005linuxdevicedrivers}. Kernel consumers allocate PDs, CQs, QPs, and register memory regions. Kernel space ownership is required when the memory being registered is not userspace memory.

GPU VRAM integration requires a bridge across the \texttt{struct page} boundary and uses NVIDIA peer to peer pinning APIs, with teardown constraints enforced by an asynchronous unpin callback.

\subsection{Positioning and closest neighbors}

Table 1 positions dmaplane relative to systems that occupy adjacent layers.

\begin{table*}[t]
\centering
\scriptsize
\caption{Capability positioning across the buffer orchestration and transport stack.}
\label{tab:1}
\setlength{\tabcolsep}{5pt}
\renewcommand{\arraystretch}{1.15}
\begin{tabularx}{\textwidth}{>{\raggedright\arraybackslash}X*{5}{>{\raggedright\arraybackslash}X}}
\toprule
Capability & dmaplane & nvidia-peermem & NCCL & Mooncake & TransferEngine \\
\midrule
Buffer allocation (coherent, page-backed) & Yes & No & internal & No & No \\
NUMA-aware placement & Yes & No & topology detection & mentioned & mentioned \\
dma-buf export or import & Yes & importer only & No & No & No \\
MR registration (kernel-space) & Yes & Yes (GPU BAR only) & internal & via libibverbs & via libibverbs \\
Credit-based flow control & Yes & No & per-channel credits & implicit in batching & per-connection \\
GPU BAR pinning and mapping & Yes & Yes & via CUDA & via CUDA & via CUDA \\
Cross-machine RDMA transport & loopback and peer QP & No & multi-transport & multi-transport & multi-transport \\
Kernel tracepoint instrumentation & Yes & No & limited & No & No \\
Collective operations (AllReduce) & No & No & Yes & No & No \\
Serving integration (batching, routing) & No & No & No & Yes & No \\
\bottomrule
\end{tabularx}
\normalsize
\end{table*}

\textbf{NCCL.} NCCL~\cite{nccl} implements GPU optimized collective communication and includes internal buffer pool and topology logic. It targets barrier synchronized collectives. dmaplane targets point to point pipelines and provides the kernel level buffer lifecycle and completion safety substrate beneath collective scheduling.

\textbf{Mooncake and TransferEngine.} Mooncake~\cite{qin2024mooncake} and TransferEngine~\cite{licker2025rdmapointtopointcommunicationllm} provide transport abstractions across memory tiers and network providers in userspace. They assume that pinned memory, registration, and placement preconditions are satisfied. dmaplane externalizes and measures those preconditions by implementing kernel side allocation, sharing, and registration steps.

\textbf{nvidia-peermem.} nvidia-peermem~\cite{nvidiapeermem} bridges GPU BAR pages into the RDMA stack by registering them as memory regions. It addresses a narrow GPU to NIC sharing path. dmaplane incorporates an equivalent pinning and teardown discipline as one subsystem and extends it with NUMA placement, dma-buf export, flow control, and observability.

\section{Architecture and Design Invariants}

dmaplane is a loadable Linux kernel module that registers one character device and exposes a stable UAPI via \texttt{/dev/dmaplane}. The device is a kernel service, not a PCIe device driver. It provides buffer orchestration primitives that other components can consume.
\subsection{Architecture overview}

Figure 1 shows the system structure. Userspace submits work through \texttt{ioctl} and maps buffers via \texttt{mmap}. Work executes inside per channel worker threads draining submission rings and posting completions to completion rings. Buffers are named objects referenced by IDs, allowing subsystems to compose without exposing raw pointers across the UAPI.

For receiver notification with remote placement, dmaplane uses RDMA WRITE WITH IMMEDIATE as illustrated in Section 5. Each operation produces a send completion on the sender and a receive completion on the receiver carrying a 32 bit immediate value. Each operation consumes one pre posted receive WR on the receiver.

\FloatBarrier
\begin{figure}[t]
\centering
\includegraphics[width=\columnwidth]{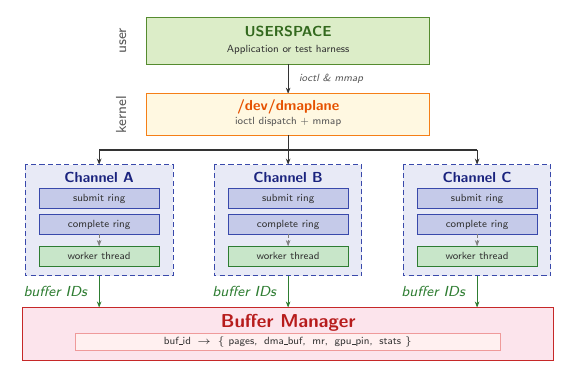}
\caption{dmaplane block diagram.}
\Description{dmaplane block diagram.}
\label{fig:1}
\end{figure}

The complete ioctl map, benchmark commands, and test mapping are provided in the appendices.
\subsection{Locking model}

dmaplane is concurrent. Multiple userspace threads can issue ioctls concurrently, and each channel has an independent worker thread. The module uses a strict lock ordering:

\begin{itemize}
\item \texttt{dev\_mutex} serializes global device operations that can sleep.

\item \texttt{rdma\_sem} is a reader writer semaphore protecting the RDMA context, taking write mode for setup and teardown and read mode for fast paths.

\item \texttt{buf\_lock} protects the buffer ID map and fast per buffer state transitions.

\item \texttt{mr\_lock} protects the MR registry.
\end{itemize}

The ordering is \texttt{dev\_mutex}, then \texttt{rdma\_sem}, then \texttt{buf\_lock}, then \texttt{mr\_lock}.
\subsection{Design invariants}

The subsystems are coupled by invariants for lifetime, teardown, and safety.

\begin{itemize}
\item \textbf{Lock ordering invariant.} Locks are acquired in one global order: \texttt{dev\_mutex}, then \texttt{rdma\_sem}, then \texttt{buf\_lock}, then \texttt{mr\_lock}. Enforcement: code paths follow the order and do not invert it. Failure prevented: deadlock under concurrent ioctls.

\item \textbf{Teardown ordering invariant.} debugfs entries are removed before device teardown, and RDMA teardown excludes in flight operations by taking \texttt{rdma\_sem} in write mode. Enforcement: module exit order and semaphore usage. Failure prevented: use after free from debugfs reads or completion processing racing teardown.

\item \textbf{mmap lifetime invariant.} A buffer cannot be destroyed while it has active userspace mappings. Enforcement: \texttt{mmap\_count} accounting on initial \texttt{mmap} plus VMA open and close callbacks. Failure prevented: freeing pages still mapped in a process VMA.

\item \textbf{dma-buf exporter invariant.} Scatter gather tables are constructed per importer attachment. Enforcement: dma-buf exporter \texttt{map\_dma\_buf} builds an SG table for the importing device. Failure prevented: invalid DMA addresses from wrong IOMMU context.

\item \textbf{GPU unpin callback invariant.} The GPU unpin callback must not block and must not run teardown under foreign locks. Enforcement: atomic revocation flag and deferred cleanup. Failure prevented: deadlock with NVIDIA driver internal locks.

\item \textbf{Completion processing invariant.} Completion polling is performed in controlled contexts and quiesced during teardown. Enforcement: polling design and teardown ordering. Failure prevented: asynchronous callbacks executing after resources are freed.
\end{itemize}

\section{Core Mechanisms}

This section presents the mechanisms that implement the buffer orchestration contract: command channels, buffer lifecycle and sharing, kernel RDMA ownership, completion safe flow control, and GPU integration constraints.
\subsection{Channels and ring based dispatch}

Each channel provides a submission ring and completion ring plus a worker thread. Userspace submits an entry, the worker executes the operation, and a completion entry returns status and metadata. Rings are fixed size circular buffers with head and tail indices protected by per ring spinlocks. Worker threads sleep on wait queues and wake on submission, and they stop via \texttt{kthread\_stop} during teardown.

This mechanism provides a stable execution substrate for later subsystems, where the dominant costs come from DMA mapping, RDMA posting, or GPU BAR access rather than ring dispatch.
\subsection{Buffer lifecycle, mmap, and dma-buf export}

dmaplane supports two allocation paths:

\begin{itemize}
\item Coherent allocation via \texttt{dma\_alloc\_coherent} for small control structures.

\item Page backed allocation via \texttt{alloc\_pages} or \texttt{alloc\_pages\_\allowbreak node} for large data buffers that require NUMA steering, scatter gather, and MR registration.
\end{itemize}

Buffers are exposed to userspace without copies by mapping the backing pages into a VMA via \texttt{vm\_insert\_page} in a loop. Buffer teardown must prevent freeing memory while it remains mapped. dmaplane tracks \texttt{mmap\_count} and rejects destroy if the count is non zero. A kernel detail matters: the VMA open callback does not run on the initial \texttt{mmap} call, so the initial mapping increments the count explicitly.

For cross device sharing, dmaplane exports buffers via dma-buf. The exporter provides callbacks to attach and map the buffer for each importing device. A key constraint is per importer scatter gather construction because DMA addresses depend on the importing device and its IOMMU context.

\begin{figure}[t]
\centering
\includegraphics[width=\columnwidth]{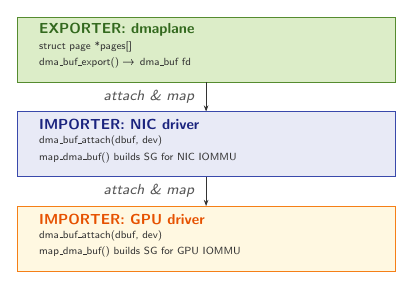}
\caption{dma-buf export and per importer attachment mapping.}
\Description{dma-buf export and per importer attachment mapping.}
\label{fig:2}
\end{figure}

\subsection{Kernel RDMA engine ownership}

dmaplane owns RDMA resources in kernel space because it owns the pages being registered. Userspace verbs APIs register userspace pages, not kernel allocated pages.

The verbs resource chain is PD, CQs, QPs, and MRs. dmaplane allocates these in order and destroys them in reverse order, flushing QPs by transitioning to error state before destroy. \texttt{IB\_POLL\_DIRECT} is used for CQs so completions are polled explicitly rather than delivered via asynchronous callbacks.

\begin{figure}[t]
\centering
\includegraphics[width=\columnwidth]{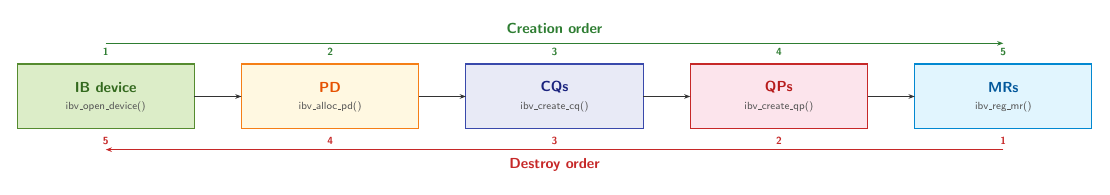}
\caption{RDMA resource hierarchy.}
\Description{RDMA resource hierarchy.}
\label{fig:3}
\end{figure}

\subsection{Completion safe flow control}

Completion queue overflow can discard completions and corrupt sender accounting. dmaplane enforces a credit invariant that bounds in flight operations by CQ capacity. Credits decrement on post and increment on completion poll, ensuring \texttt{in\_flight <= max\_credits <= CQ\_depth} in the configured benchmark paths.

RDMA WRITE WITH IMMEDIATE also consumes receiver side receive WR capacity. A receiver pre posted receive window therefore acts as a second credit type. Safe operation requires bounding in flight WRITE WITH IMMEDIATE operations by both sender completion capacity and receiver notification capacity.

\begin{figure}[t]
\centering
\includegraphics[width=\columnwidth]{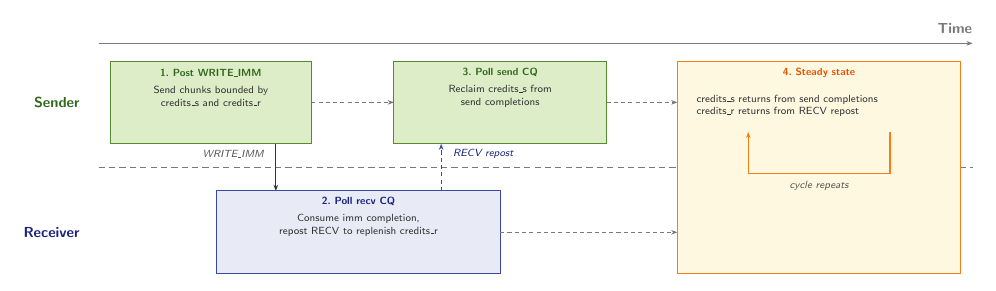}
\caption{Timeline of send CQ credits and receive window credits in a WRITE WITH IMMEDIATE pipeline.}
\Description{Timeline diagram showing a sender posting WRITE WITH IMMEDIATE chunks bounded by send credits and receive window credits, the receiver polling the receive completion queue and reposting receives, and the sender polling the send completion queue to reclaim credits.}
\label{fig:4}
\end{figure}

\subsection{GPU memory integration constraints}

GPU VRAM does not provide \texttt{struct page}, so standard kernel DMA and RDMA mapping paths do not apply directly. dmaplane uses NVIDIA peer to peer pinning APIs to obtain BAR page descriptors and enforces a non blocking unpin callback contract by deferring cleanup outside the callback.

Host access to BAR mappings is sensitive to mapping type. Write combining mappings improve host to GPU write throughput relative to uncacheable mappings, while GPU to host reads remain limited by non posted PCIe read semantics. The measured tier behavior is reported in Table 5.

\section{Disaggregated Inference Demonstration}

This section demonstrates composition of the buffer orchestration layer by transferring KV cache state between two machines and running a decode phase on the receiver.
\subsection{Pipeline overview}

The demo separates inference into two roles:

\begin{itemize}
\item \textbf{Prefill machine.} Runs tokenization and the forward pass to produce KV cache. Consolidates KV cache into a pinned GPU staging buffer and transfers it in fixed size chunks to the receiver.

\item \textbf{Decode machine.} Pre posts receives, accepts RDMA WRITE WITH IMMEDIATE notifications, reconstructs KV cache tensor views over the landing zone, and runs token generation.
\end{itemize}

The demo uses RDMA WRITE WITH IMMEDIATE so that each chunk is placed into a specific remote address range and also delivers a 32 bit immediate value identifying the chunk.

\subsection{Protocol and flow control}

Each KV cache transfer is chunked. For each chunk the sender posts \texttt{IB\_WR\_RDMA\_WRITE\_WITH\_IMM} with:

\begin{itemize}
\item local SGE referencing the staged chunk

\item remote address within the receiver landing zone MR

\item receiver rkey

\item immediate value encoding of \texttt{layer\_index} and \texttt{chunk\_index} (16-bit fields) as implemented in the accompanying artifact
\end{itemize}

On the receiver, each WRITE WITH IMMEDIATE consumes one pre posted receive WR and produces a receive completion carrying the immediate value. The receive window therefore provides receiver notification credits as defined in Section 4.4, and the sender side flow control provides send CQ credits. The combined bound applies because the verb produces completions on both sides.

A sentinel immediate value signals completion of the transfer, and the receiver verifies that expected chunks arrived before reconstruction.
\subsection{Measured timing breakdown}

Table 2 reports the measured pipeline timing breakdown for the two instance EC2 demo using Soft-RoCE.

\begin{table}[t]
\centering
\scriptsize
\caption{Disaggregated inference pipeline timing (two EC2 `g5.xlarge`, Soft-RoCE).}
\label{tab:2}
\setlength{\tabcolsep}{3pt}
\renewcommand{\arraystretch}{1.15}
\begin{tabularx}{\columnwidth}{>{\raggedright\arraybackslash}p{0.68\columnwidth} >{\raggedleft\arraybackslash}p{0.22\columnwidth}}
\toprule
Phase & Time \\
\midrule
Tokenization & 1.2 ms \\
Prefill forward pass & 45.3 ms \\
KV-cache consolidation & 0.8 ms \\
KV-cache transfer & 52.1 ms \\
KV-cache reconstruction & 0.003 ms \\
--------------------------- & -------- \\
Time-to-first-token (TTFT) & 98.2 ms \\
--------------------------- & -------- \\
Decode throughput & 45.3 tok/s \\
Decode latency (per token) & 22 ms average \\
\bottomrule
\end{tabularx}
\normalsize
\end{table}

\section{Evaluation and Discussion}

This section presents four result sets that support the buffer orchestration thesis: flow control safety under sustained RDMA load (Table 3), NUMA placement sensitivity at DRAM scale (Table 4), GPU BAR tier behavior versus cudaMemcpy (Table 5), and the end to end demo timing breakdown (Table 2). All other microbenchmarks, extended tables, and setup details are provided in the appendices.

\subsection{Flow control under sustained streaming}

Table 3 reports sustained streaming and stress results under Soft-RoCE loopback. The reported configurations include a high credit configuration and a low credit stress configuration that generates large numbers of stalls while reporting zero CQ overflows.

\begin{table}[t]
\centering
\scriptsize
\caption{RDMA streaming and flow control results (Soft-RoCE loopback).}
\label{tab:3}
\setlength{\tabcolsep}{3pt}
\renewcommand{\arraystretch}{1.15}
\begin{tabularx}{\columnwidth}{>{\raggedright\arraybackslash}p{0.66\columnwidth} >{\raggedleft\arraybackslash}p{0.24\columnwidth}}
\toprule
Metric & Value \\
\midrule
Sustained throughput (10 s, max\_credits=64) & 1,037 MB/s average \\
Per-second window range & 1,047 to 1,087 MB/s \\
Window spread & 3.8\% \\
CQ overflows & 0 \\
Credit stalls & 0 \\
Stress configuration (max\_credits=4, high=3, low=1) &  \\
Credit stalls (5 s) & 72.7 million \\
CQ overflows & 0 \\
\bottomrule
\end{tabularx}
\normalsize
\end{table}

These results are consistent with the invariant \texttt{in\_flight <= max\_credits <= CQ\_depth} for the signaled work request configuration used in the evaluation.
\subsection{NUMA placement sensitivity}

Table 4 reports cross node CPU memcpy bandwidth on AWS \texttt{c5.metal} at 1 MB and 64 MB. The 1 MB case does not show a cross node penalty, while the 64 MB case does, consistent with cache hiding effects at small working sets.

\begin{table*}[t]
\centering
\scriptsize
\caption{Cross node memcpy bandwidth on AWS `c5.metal`. Values are MB/s.}
\label{tab:4}
\setlength{\tabcolsep}{5pt}
\renewcommand{\arraystretch}{1.15}
\begin{tabularx}{\textwidth}{>{\centering\arraybackslash}p{0.10\textwidth}*{4}{>{\centering\arraybackslash}p{0.12\textwidth}}>{\raggedright\arraybackslash}X}
\toprule
Buffer size & Node 0 to 0 & Node 0 to 1 & Node 1 to 0 & Node 1 to 1 & Cross-node penalty \\
\midrule
1 MB & 13,284 & 13,218 & 12,976 & 13,051 & < 1\% (fits in cache) \\
64 MB & 6,778 & 5,577 & 5,013 & 6,095 & 18\% symmetric \\
\bottomrule
\end{tabularx}
\normalsize
\end{table*}

Methodology: based on the accompanying artifact; iteration count and warmup policy are not specified in the paper.

The practical implication for buffer orchestration is that placement errors can be silent and can appear only at DRAM scale buffer sizes.

\subsection{GPU memory access tiers}

Table 5 consolidates measured tiers for GPU BAR access and cudaMemcpy on RTX 5000 Ada, plus measured software RDMA loopback and cross machine behavior from the accompanying artifact.

\begin{table*}[t]
\centering
\scriptsize
\caption{GPU memory access hierarchy (RTX 5000 Ada). Cross-machine values are over 1 Gigabit Ethernet to a remote NUC.}
\label{tab:5}
\setlength{\tabcolsep}{5pt}
\renewcommand{\arraystretch}{1.15}
\begin{tabularx}{\textwidth}{>{\raggedright\arraybackslash}X>{\centering\arraybackslash}p{0.16\textwidth}>{\centering\arraybackslash}p{0.16\textwidth}>{\raggedright\arraybackslash}p{0.28\textwidth}}
\toprule
Access method & Host to GPU (MB/s) & GPU to host (MB/s) & Section \\
\midrule
UC BAR (ioremap) & 44 & 6 & Accompanying artifact \\
WC BAR (ioremap\_wc, 4 MB) & 10,097 & 107 & Accompanying artifact \\
cudaMemcpy (16 MB) & 12,552 & 13,124 & Accompanying artifact \\
GPU RDMA loopback (rxe, 64 KB) & n/a & \textasciitilde{}20 & Accompanying artifact \\
GPU RDMA cross-machine (1 Gigabit) & n/a & 11.6 & Accompanying artifact \\
\bottomrule
\end{tabularx}
\normalsize
\end{table*}

\noindent\textit{Note on GPU RDMA loopback entry:} The \textasciitilde{}20~MB/s effective rate is derived from the measured 64~KB loopback latency of 3.2~ms (64~KB / 3.2~ms $\approx$ 20~MB/s).

Methodology: BAR and cudaMemcpy tiers are mean of three runs per size; cross machine tiers are from the two machine configuration described in the paper.

The NIC DMA engine BAR read tiers described in NVIDIA documentation are not measured in this work and must be treated as external claims when referenced.

\subsection{End to end demo}

Table 2 provides the end to end timing breakdown for the disaggregated inference demo. The largest component in time to first token in the reported run is the KV cache transfer time under Soft-RoCE and Ethernet.
\subsection{Limitations and generality}

\subsubsection{Measured on Soft-RoCE}

RDMA measurements in this paper run over Soft-RoCE, a software provider that executes verbs in kernel software and uses CPU memcpy. Results reflect this provider behavior and host CPU scheduling effects.
\subsubsection{Provider independent by construction}

dmaplane uses provider independent kernel verbs APIs and standard completion semantics. The flow control invariant is expressed over completion accounting rules, and the dma-buf sharing model follows the kernel framework contract. These properties do not depend on a specific RDMA NIC provider.
\subsubsection{Not validated on hardware RDMA}

The paper does not include performance measurements on hardware RDMA NICs such as ConnectX or EFA. GPU pinning uses the same peer to peer API used in GPUDirect RDMA environments~\cite{gpudirectrdma}, but NIC BAR DMA tiers are not measured in this work.

Implementation details, scripts, and additional measurements are provided in the accompanying artifact and appendices.

\section{Conclusion}

This paper presents dmaplane, a Linux kernel module that implements buffer orchestration as a first class layer beneath transport and above devices. dmaplane provides a stable kernel UAPI and composes mechanisms for buffer lifecycle, cross device sharing, kernel RDMA ownership, completion safe flow control, NUMA aware placement, observability, and GPU BAR pinning constraints.

Four measured result sets support the framing:

\begin{itemize}
\item NUMA placement penalties can be hidden at cache scale and appear at DRAM scale (Table 4).

\item Credit based flow control prevents CQ overflow under sustained load in the reported configurations (Table 3).

\item BAR mapping tier choice changes host to GPU throughput dramatically while GPU to host reads remain limited (Table 5).

\item The end to end demo composes these primitives to transfer KV cache and run decode on the receiver (Table 2).
\end{itemize}

Future work includes validation on hardware RDMA providers and integration paths where dmaplane is consumed by higher level transports or serving frameworks without changing the kernel layer contract.

\nocite{*}
\bibliographystyle{ACM-Reference-Format}
\bibliography{references}

@inproceedings{patel2024splitwise,
  title        = {Splitwise: Efficient Generative LLM Inference Using Phase Splitting},
  author       = {Patel, P. and Choukse, E. and Zhang, C. and Goiri, I. and Shah, A. and Maleki, S. and Bianchini, R.},
  booktitle    = {Proceedings of the 51st International Symposium on Computer Architecture (ISCA)},
  year         = {2024}
}

@inproceedings{zhong2024distserve,
  title        = {DistServe: Disaggregating Prefill and Decoding for Goodput-optimized Large Language Model Serving},
  author       = {Zhong, Y. and Liu, S. and Chen, J. and Hu, J. and Zhu, Y. and Liu, X. and Jin, X. and Zhang, H.},
  booktitle    = {Proceedings of the 18th USENIX Symposium on Operating Systems Design and Implementation (OSDI)},
  year         = {2024}
}

@misc{qin2024mooncake,
  title        = {Mooncake: A KVCache-centric Disaggregated Architecture for LLM Serving},
  author       = {Qin, R. and Li, Z. and He, W. and Zhang, M. and Qiu, Y. and Wang, L. and Lu, J.},
  year         = {2024}
}

@inproceedings{kwon2023pagedattention,
  title        = {Efficient Memory Management for Large Language Model Serving with PagedAttention},
  author       = {Kwon, W. and Li, Z. and Zhuang, S. and Sheng, Y. and Zheng, L. and Yu, C. and Gonzalez, J. and Zhang, H. and Stoica, I.},
  booktitle    = {Proceedings of the 29th ACM Symposium on Operating Systems Principles (SOSP)},
  year         = {2023}
}

@misc{zheng2023sglang,
  title        = {SGLang: Efficient Execution of Structured Language Model Programs},
  author       = {Zheng, L. and Yin, L. and Xie, Z. and Huang, J. and Sun, C. and Yu, C. and Cao, S. and Kozyrakis, C. and Stoica, I. and Gonzalez, J. and Barrett, C. and Zhang, H.},
  year         = {2023}
}

@misc{licker2025rdmapointtopointcommunicationllm,
  title         = {RDMA Point-to-Point Communication for LLM Systems},
  author        = {Licker, Nandor and Hu, Kevin and Zaytsev, Vladimir and Chen, Lequn},
  year          = {2025},
  eprint        = {2510.27656},
  archivePrefix = {arXiv},
  primaryClass  = {cs.DC},
  url           = {https://arxiv.org/abs/2510.27656}
}

@misc{nccl,
  title        = {NCCL: NVIDIA Collective Communications Library},
  author       = {{NVIDIA Corporation}},
  howpublished = {Documentation},
  url          = {https://docs.nvidia.com/deeplearning/nccl/},
  year         = {2026}
}

@book{corbet2005linuxdevicedrivers,
  title     = {Linux Device Drivers},
  author    = {Corbet, Jonathan and Rubini, Alessandro and Kroah-Hartman, Greg},
  edition   = {3},
  publisher = {O'Reilly Media},
  year      = {2005}
}

@inproceedings{pfaff2015ovs,
  title     = {The Design and Implementation of Open vSwitch},
  author    = {Pfaff, Ben and Pettit, Justin and Koponen, Teemu and Jackson, Ethan J. and Zhou, Andy and Rajahalme, Jarno and Gross, Justin and Wang, Alex and Stringer, Jeremy and Shelar, Pravin and Amidon, Keith and Casado, Martin},
  booktitle = {Proceedings of the 12th USENIX Symposium on Networked Systems Design and Implementation (NSDI)},
  year      = {2015}
}

@misc{gpudirectrdma,
  title        = {GPUDirect RDMA},
  author       = {{NVIDIA Corporation}},
  howpublished = {Documentation},
  url          = {https://docs.nvidia.com/cuda/gpudirect-rdma/},
  year         = {2026}
}

@misc{nvidiapeermem,
  title        = {nvidia-peermem: NVIDIA Peer Memory Client},
  author       = {{NVIDIA Corporation}},
  howpublished = {Kernel module documentation},
  year         = {2026}
}

\appendix
\section{Methodology}

This appendix provides (1) a phase-to-subsystem mapping retained as internal project terminology and (2) a development process note describing AI-assisted development.

\subsection{Development process note}

The phase structure was used to bound implementation scope, define interfaces, and isolate test surfaces. dmaplane was built using Claude Code as a collaborative tool for implementation, debugging, and test suite generation. The bounded scope of each phase, with a clear specification, a defined interface to previous phases, and an isolated test surface, supported iterative development and verification.

\section{Build and Deployment}

This appendix consolidates build and deployment details referenced throughout the paper. The repo source code is at https://github.com/marcoeg/dmaplane

\subsection{Kernel and configuration requirements}

The work targets a stock Ubuntu kernel (6.5). Required kernel facilities include:

\begin{itemize}
\item RDMA core (\texttt{ib\_core}) and the kernel verbs interface

\item Soft-RoCE provider (\texttt{rdma\_\allowbreak rxe}) for the software provider measurements

\item debugfs (\texttt{CONFIG\_DEBUG\_FS}) for debugfs inspection

\item ftrace and tracepoints for optional runtime tracing
\end{itemize}

If debugfs is not mounted or \texttt{CONFIG\_DEBUG\_FS} is disabled, debugfs initialization returns silently and is not treated as fatal.

\subsection{Module build and load}

The module is built out of tree using Kbuild,  the Linux kernel build sytstem. When tracepoints are enabled, the build must satisfy \texttt{TRACE\_INCLUDE\_PATH} and include path requirements so \texttt{define\_trace.h} can re-include the trace header during macro expansion.

Loading uses standard module tooling:
\begin{Verbatim}[fontsize=\footnotesize,breaklines,breakanywhere]
sudo insmod dmaplane.ko
ls -l /dev/dmaplane
\end{Verbatim}

Unloading requires that userspace closes the device file and releases mappings so teardown invariants are satisfied:
\begin{Verbatim}[fontsize=\footnotesize,breaklines,breakanywhere]
sudo rmmod dmaplane
\end{Verbatim}

\subsection{Soft-RoCE bring-up (rxe)}

Soft-RoCE is used for the RDMA provider in the measurements. A typical bring-up sequence:
\begin{Verbatim}[fontsize=\footnotesize,breaklines,breakanywhere]
sudo modprobe rdma_rxe
sudo rdma link add rxe_eth0 type rxe netdev <netdev>
ls /sys/class/infiniband
\end{Verbatim}

The RDMA device name (for example, \texttt{rxe\_eth0}) is passed to dmaplane through the RDMA setup ioctl.

\subsection{debugfs mount and inspection}

Mount debugfs if not already mounted:
\begin{Verbatim}[fontsize=\footnotesize,breaklines,breakanywhere]
sudo mount -t debugfs none /sys/kernel/debug
\end{Verbatim}

Inspect driver state:
\begin{Verbatim}[fontsize=\footnotesize,breaklines,breakanywhere]
cat /sys/kernel/debug/dmaplane/stats
cat /sys/kernel/debug/dmaplane/buffers
cat /sys/kernel/debug/dmaplane/rdma
cat /sys/kernel/debug/dmaplane/flow
cat /sys/kernel/debug/dmaplane/histogram
\end{Verbatim}

\subsection{Tracepoint enablement}

Enable tracepoints via ftrace:
\begin{Verbatim}[fontsize=\footnotesize,breaklines,breakanywhere]
echo 1 > /sys/kernel/debug/tracing/events/dmaplane/dmaplane_rdma_post/enable
echo 1 > /sys/kernel/debug/tracing/events/dmaplane/dmaplane_rdma_completion/enable
cat /sys/kernel/debug/tracing/trace_pipe
\end{Verbatim}

\subsection{NVIDIA driver constraints and runtime symbol bridge}

GPU integration requires the NVIDIA kernel driver and the NVIDIA peer to peer API symbols. Because dmaplane imports GPL only symbols from \texttt{ib\_core}, it does not take a static link dependency on \texttt{nvidia.ko}. Instead, it resolves symbols at runtime using \texttt{symbol\_get} and releases them using \texttt{symbol\_put}. If symbols are unavailable, GPU related ioctls return \texttt{-ENODEV} and GPU features are disabled.
\section{Test and Observability Map}

This appendix summarizes the test surface and observability mechanisms described in the paper and referenced by the main body. Where specific filenames are not provided in the accompanying artifact, entries are marked as not specified.
\subsection{Test suite map}

\begin{table*}[t]
\centering
\scriptsize
\caption{Test suite map by subsystem.}
\label{tab:c1}
\setlength{\tabcolsep}{6pt}
\renewcommand{\arraystretch}{1.15}
\begin{tabularx}{\textwidth}{>{\raggedright\arraybackslash}p{0.22\textwidth} >{\raggedright\arraybackslash}p{0.20\textwidth} >{\raggedright\arraybackslash}X}
\toprule
Subsystem & Test entry point & Pass condition \\
\midrule
Device and ioctl dispatch & not specified & Module loads, \texttt{/dev/dmaplane} present, ioctls return expected codes. \\
Rings and workers & userspace stress harness & No kernel warnings, no lockdep issues, no data corruption, clean unload. \\
DMA buffers and mmap & not specified & Create buffer, mmap, write and verify, munmap, destroy succeeds. \\
dma-buf export & not specified & Export returns dma-buf fd, attach and detach paths do not leak, release callback fires. \\
RDMA setup and SEND and RECV & ioctl benchmarks & QP setup succeeds, SEND and RECV completes, CQs polled without errors. \\
MR registration & ioctl MR tests & MR registration returns lkey and rkey when requested, deregistration releases mappings. \\
NUMA topology & \texttt{IOCTL\_QUERY\_NUMA\_\allowbreak TOPO} & Returns distance matrix and node inventory without kernel warnings. \\
NUMA memcpy benchmark & \texttt{IOCTL\_NUMA\_\allowbreak BENCH} & Produces per-cell bandwidth values without kernel warnings. \\
Flow control & \texttt{IOCTL\_SUSTAINED\_\allowbreak STREAM} & Zero CQ overflows under configurations in Table 3. \\
QD sweep & \texttt{IOCTL\_QD\_\allowbreak SWEEP} & Produces throughput and latency values (Table 4). \\
Instrumentation & debugfs inspection & debugfs files readable when enabled, histogram format produced during streaming. \\
GPU pin and BAR mapping & \texttt{IOCTL\_GPU\_\allowbreak PIN} and \texttt{IOCTL\_GPU\_\allowbreak BENCH} & Pin succeeds when NVIDIA driver present, unpin callback safe. \\
GPU RDMA loopback & not specified & Loopback transfer verifies bytes match, no kernel warnings. \\
WRITE WITH IMMEDIATE helpers & \texttt{IOCTL\_RDMA\_\allowbreak WRITE\_\allowbreak IMM} and related ioctls & Receiver polls \texttt{imm\_data} and receives expected identifiers. \\
Disaggregated inference & prefill and decode scripts & Demo produces coherent output and timings consistent with Table 2. \\
\bottomrule
\end{tabularx}
\end{table*}

Methodology: derived from the paper text and artifact descriptions; specific script and binary names are not fully specified in the paper.
\subsection{Observability summary}

dmaplane provides two observability paths:

\begin{itemize}
\item \textbf{Tracepoints.} Optional kernel tracepoints for events such as RDMA posting, completions, and flow stalls. When disabled, tracepoints compile to near no op behavior.

\item \textbf{debugfs.} Read-only debugfs files under \path{/sys/kernel/debug/dmaplane/} support live inspection of counters, buffers, RDMA state, flow state, and the latency histogram. The histogram output format is described in this paper.
\end{itemize}

\section{Glossary}
{\scriptsize\raggedright\sloppy
\begin{description}[leftmargin=1.6em,style=nextline]
\item[ACPI] Advanced Configuration and Power Interface. Firmware interface that provides tables used by the OS, including NUMA topology.
\item[BAR] Base Address Register. PCIe mechanism that maps a device memory region into the host physical address space.
\item[BAR1] NVIDIA term for the PCIe BAR aperture that exposes portions of GPU VRAM to the host.
\item[CAS] Compare and swap. Atomic primitive used to update shared values without locks.
\item[CBQ] Completion queue. See CQ.
\item[CQ] Completion Queue. RDMA data structure that receives Work Completions for completed operations.
\item[CQE] Completion Queue Entry. A single completion record placed into a CQ.
\item[CPU] Central Processing Unit.
\item[CUDA] NVIDIA Compute Unified Device Architecture. Platform and programming model for NVIDIA GPUs.
\item[DMA] Direct Memory Access. Mechanism where a device reads or writes memory without CPU copying.
\item[DRAM] Dynamic Random Access Memory. System main memory.
\item[EFA] Elastic Fabric Adapter. AWS network interface supporting high performance communication, including RDMA like semantics.
\item[ENA] Elastic Network Adapter. AWS virtual network interface used on many EC2 instances.
\item[GID] Global Identifier. RDMA addressing identifier used in RoCE and InfiniBand.
\item[GPU] Graphics Processing Unit.
\item[HCA] Host Channel Adapter. RDMA capable network adapter used in InfiniBand and RoCE environments.
\item[HBM] High Bandwidth Memory. GPU attached memory (for example HBM2e) with very high bandwidth relative to DRAM.
\item[IB] InfiniBand. RDMA interconnect architecture and related kernel subsystem interfaces.
\item[IOMMU] Input Output Memory Management Unit. Hardware that translates device DMA addresses to physical memory and enforces access control.
\item[KV] cache   Key value cache. Transformer attention state produced during prefill and consumed during decode.
\item[LID] Local Identifier. InfiniBand subnet local address.
\item[LLC] Last Level Cache. Typically the largest shared CPU cache (often L3).
\item[LLM] Large Language Model.
\item[MR] Memory Region. RDMA object authorizing NIC access to a range of memory, identified by lkey and rkey.
\item[NIC] Network Interface Card.
\item[NUMA] Non Uniform Memory Access. System architecture where memory is divided into nodes with different access costs depending on CPU locality.
\item[P2P] Peer to peer. In this paper, typically refers to GPU peer to peer memory pinning and access for third party devices.
\item[PD] Protection Domain. RDMA resource that scopes access permissions for QPs, CQs, and MRs.
\item[PFC] Priority Flow Control. Ethernet mechanism used to reduce packet loss in RoCE deployments.
\item[QP] Queue Pair. RDMA endpoint containing a Send Queue and Receive Queue.
\item[RC] Reliable Connected. RDMA transport type that provides reliable in order delivery.
\item[RDMA] Remote Direct Memory Access. Mechanism enabling one host to read or write memory on another host without remote CPU involvement in the data path.
\item[RECV] WR    Receive Work Request. RDMA receive queue entry posted by the receiver to accept incoming SEND or WRITE WITH IMMEDIATE notifications.
\item[SG] Scatter gather. Representation of a buffer as a list of segments, often page sized, used for DMA mapping.
\item[SGE] Scatter Gather Element. One segment descriptor referenced by an RDMA work request.
\item[SGT] Scatter Gather Table. Collection of SG entries describing a buffer.
\item[SLIT] System Locality Information Table. ACPI table describing relative NUMA node distance.
\item[SQ] Send Queue. RDMA queue inside a QP that holds outgoing work requests.
\item[SOSP] ACM Symposium on Operating Systems Principles.
\item[TLP] Transaction Layer Packet. PCIe transaction unit for reads and writes.
\item[TTFT] Time to first token. Time from request start to the first generated token in an inference pipeline.
\item[UPI] Ultra Path Interconnect. Intel inter socket interconnect.
\item[VMA] Virtual Memory Area. Kernel structure representing a contiguous memory mapping in a process.
\item[VRAM] Video RAM. GPU local memory.
\item[WC] Write combining. CPU memory mapping type that buffers and coalesces writes, used for BAR mappings via ioremap\_wc.
\item[WR] Work Request. RDMA operation descriptor posted to a QP, such as SEND or RDMA WRITE.
\end{description}
}
\FloatBarrier
\end{document}